# Convolutional neural networks applied to differential dynamic microscopy reduces noise when quantifying heterogeneous dynamics

Gildardo Martinez, Justin Siu, Steven Dang, Dylan Gage, Emma Kao, Juan Carlos Avila, Ruilin You, and Ryan McGorty*

*Department of Physics and Biophysics, University of San Diego, San Diego, CA 92110, USA*

**Abstract**: Differential dynamic microscopy (DDM) typically relies on movies containing hundreds or thousands of frames to accurately quantify motion in soft matter systems. Using movies much shorter in duration produces noisier and less accurate results. This limits the applicability of DDM to situations where the dynamics are stationary over extended times. Here, we investigate a method to denoise the DDM process, particularly suited to when a limited number of imaging frames are available or when dynamics are quickly evolving in time. We use a convolutional neural network encoder-decoder (CNN-ED) model to reduce the noise in the intermediate scattering function that is computed via DDM. We demonstrate this approach of combining machine learning and DDM on samples containing diffusing micron-sized colloidal particles. We quantify how the particles' diffusivities change over time as the fluid they are suspended in gels. We also quantify how the diffusivity of particles varies with position in a sample containing a viscosity gradient. These test cases demonstrate how studies of non-equilibrium dynamics and high-throughput screens could benefit from a method to denoise the outputs of DDM.

## 1 Introduction

Researchers interested in quantifying the dynamics of colloidal suspensions, gels, biological systems, or other soft materials have numerous tools to choose from, e.g., dynamic light scattering, single-particle tracking, particle image velocity, and more. A relative newcomer to the suite of available methods is differential dynamic microscopy (DDM). First introduced in 2008,[1] this technique has now been used to measure the dynamics of diffusing colloids,[1,2] active bacteria,[3] intracellular transport,[4] colloidal gels,[5,6] active networks,[7] and more. The range of systems probed with DDM will likely only expand as software packages for DDM become more robust and user friendly.

Multiple features of DDM make it an ideal tool for soft matter applications (see the many review papers[8-11] that clearly enumerate these features), but a drawback can be the need to acquire 100s or 1000s of images for the analysis. Collecting such videos can take 10s or 100s of seconds. In many scenarios, the system under investigation is, or can be well approximated to be, in a steady state during that time interval. However, in systems with evolving dynamics (e.g., a sample undergoing gelation or an active system driven out of equilibrium by some exhaustible fuel supply), there can be a difficult balance to make between acquiring images over a long enough time duration to obtain reliable statistics and having fine enough time resolution to describe the temporal evolution of the dynamics. Of course, this trade-off is in no way unique to DDM. But, as DDM relies on microscopy imaging where camera frame rates and the need for adequate signal-to-noise ratios can limit imaging rates to the 10-100 Hz range and as soft matter and biological systems often have dynamics which evolve at rates not vastly slower, DDM users in the soft matter and biophysics community may readily confront this issue.

To address this challenge, we have turned to machine learning. Machine learning approaches have demonstrated success in addressing challenges in soft matter research[12-21] and in denoising data in numerous other contexts. In X-ray or electron imaging and medical imaging situations, machine learning can counter the increased noise when using low doses of radiation to help prevent sample damage.[22,23] In fluorescence microscopy, machine learning has likewise been used to denoise images so that fluorescence excitation intensity can be minimized to reduce photodamage and photobleaching.[24-

[26] Applied to DDM, machine learning approaches have previously been used for the purposes of uncertainty quantification[27] and for help guiding DDM-based microrheology experiments.[28]

While DDM typically operates on images acquired from a microscope, our approach is not to denoise the real-space images which are then the inputs to the DDM algorithm. Rather, we denoise, using a convolutional neural network encoder-decoder (CNN-ED), one of the outputs of the DDM process: the intermediate scattering function (ISF) which we can express as a two-dimensional (2D) function of lag time and of wave vector. We show that this 2D ISF suffers from noise when we use DDM on 10s (rather than 100s or 1000s) of imaging frames, but that a trained CNN-ED model can ameliorate this situation. Our approach to this denoising process draws heavily from recent work by Konstantinova et al. [29, 30] Those authors showed that a CNN-ED model could reduce the noise in the 2D correlation functions computed for X-ray photon correlation spectroscopy.

In this work, we train and deploy a CNN-ED model with ISFs that can be well described by a single exponential function. This is typically the case with dynamics that are relatively simple such as a monodisperse colloidal suspension exhibiting Brownian diffusion. We note that DDM has previously been applied to samples showing much more complex dynamics. For example, in studies of polydisperse samples of protein clusters or colloidal particles, modified cumulant fits to the ISF were used;[31, 32] in studies of samples with multiple relaxation modes, the ISFs were fit to functions with multiple exponential terms;[11, 33-35] in studies of motile bacteria, ISFs were described by more complex functions to account for both diffusive and ballistic motion;[3, 36, 37] in studies of samples with non-ergodic dynamics, ISF models containing a non-ergodicity factor were used;[6, 38] and in studies of flowing samples, the ISFs were fit to a model to account for the flow velocities.[39] That multiple classes of dynamics can be investigated with DDM is one of the many assets of the technique. However, that asset of DDM is not explored in this work where we examine simple dynamics exhibiting a single decay mode. How machine learning could be used with DDM to study complex and multimodal dynamics and how a machine learning model trained with one class of dynamics would fare when applied to dynamics of another class are interesting questions for future studies.

In what follows, we will first expand on the principles of DDM and where in the workflow of this analysis method we insert a CNN-ED model. We then describe how we have trained a CNN-ED model using experimental images of diffusing micron-sized colloidal particles. Implementation of this model is then demonstrated on two experimental data sets. In the first experiment we observe colloidal particles suspended in a protein solution of sodium caseinate undergoing gelation. We quantify how the diffusion of these colloids slows as the system gels. Our use of the CNN-ED model to denoise data improved our ability to capture the non-stationary dynamics. Our second experimental demonstration showcases how DDM with a CNN-ED model could enable high throughput screening with an on-the-fly scanning approach. We suspend colloidal particles in a gradient of dextran and image the sample while continuously moving our imaging field of view across this gradient. These demonstrations reveal how a CNN-ED model can reduce the noise in the intermediate scattering function (and, therefore, the extracted diffusion coefficients) when employing DDM on short image sequences, as short as 51 frames in our experiments.

## 2 Background

DDM extracts scattering-like information from videos recorded using a microscope. Numerous microscopy modalities, such as bright-field, wide-field fluorescence, confocal,[40] light-sheet,[41] and dark-field,[42] have been employed, and we direct readers to the provided references for how aspects of the imaging process, such as linear space variance,[42] degree of optical section,[40, 41] multiple scattering events,[43] and coherence of the light source,[44] affect DDM analysis.

In dynamic light scattering, one measures the intensity of scattered light at one (or more) wave vectors and computes the temporal correlation of this signal to determine dynamics. With a microscope, dynamics can be extracted in an analogous manner by finding the temporal correlations of the intensities of images in Fourier space. By performing such analyses in Fourier space, one can pull out correlation functions across a range of spatial frequencies (or wave vectors) limited by the smallest resolvable element (the size of a pixel or a diffraction-limited spot) and the size of the field-of-view. From characteristic decay times of these correlation functions and how the decay times depend on the wave vector, one can determine diffusion coefficients, velocities, or other metrics of a sample's dynamics.

In detail, as presented in Fig. 1, after recording a sequence of images (Fig. 1A), $I(x, y, t)$, we compute the Fourier transform of the differences between images separated by a lag time, $\Delta t$. This is averaged over all pairs of images separated by $\Delta t$ and results in what is referred to as the DDM matrix or the image structure function (Fig 1B):

$$D(q_x, q_y, \Delta t) = \langle |\tilde{I}(q_x, q_y, t) - \tilde{I}(q_x, q_y, t + \Delta t)|^2 \rangle_t \tag{1}$$

For isotropic dynamics, this DDM matrix is azimuthally averaged to yield $D(q, \Delta t)$ where $q$ is the magnitude of the wave vector, $q = \sqrt{q_x^2 + q_y^2}$. Typically, this DDM matrix is fit to a function of the form:

$$D(q, \Delta t) = A(q)[1 - f(q, \Delta t)] + B(q) \tag{2}$$

where the $A(q)$ is an amplitude term that depends on the optical properties of the microscope and the structure of the sample, $B(q)$ is a background term that depends on the noise in the image, and $f(q, \Delta t)$ is the intermediate scattering function (ISF). For samples exhibiting diffusive dynamics, we model the ISF as an exponential function (Fig. 1C), $f(q, \Delta t) = \exp(-\Delta t/\tau(q))$ where $\tau(q)$ is a characteristic decay time related to the diffusion coefficient according to $\tau = (Dq^2)^{-1}$ (Fig. 1D). While in prior work using DDM, the ISF is typically displayed as plots of the ISF at particular values of $q$ as a function of lag time, as shown in Fig. 1C, in this work, we display the ISF as a 2D matrix (Fig. 1E) as this matrix is what we denoised using a machine learning approach.

Since DDM was first described, there have been numerous extensions to the technique. DDM has been used to perform microrheology by obtaining a mean squared displacement from $D(q, \Delta t)$.[45, 46] Strategies have been developed to mitigate the effects of objects leaving the field of view,[47] of linear space variant imaging modes,[42] of multiple scattering in turbid samples,[43] and of drift in flowing samples.[48, 49] Strategies have also been described to extend DDM's temporal resolution,[50, 51] to combine DDM with shear cells[52] or optical tweezers,[53] and to extract multiple modes of dynamics, such as both translational and rotational diffusion,[33, 54] multiple diffusivities in suspensions with multimodal size distributions,[34, 55] or both diffusive and advective motion in biological systems.[3, 4, 56] Our goal for this work was to complement DDM with machine learning to reduce noise in situations where a limited number of video frames are available for DDM analysis.

We were motivated to denoise DDM because the use of a limited number of frames in calculating the ISF can lead to imprecise or inaccurate estimates of a sample's dynamics. This is shown in Fig. 2 where the ISFs calculated from movies of varying numbers of frames, from 3,000 to 51 (corresponding to durations of 120 s to 2 s), are shown. In cases where samples exhibit steady-state dynamics, collecting a large number of frames to achieve accurate characterization is feasible. But this may not be the case for non-equilibrium samples or other situations with non-stationary dynamics.

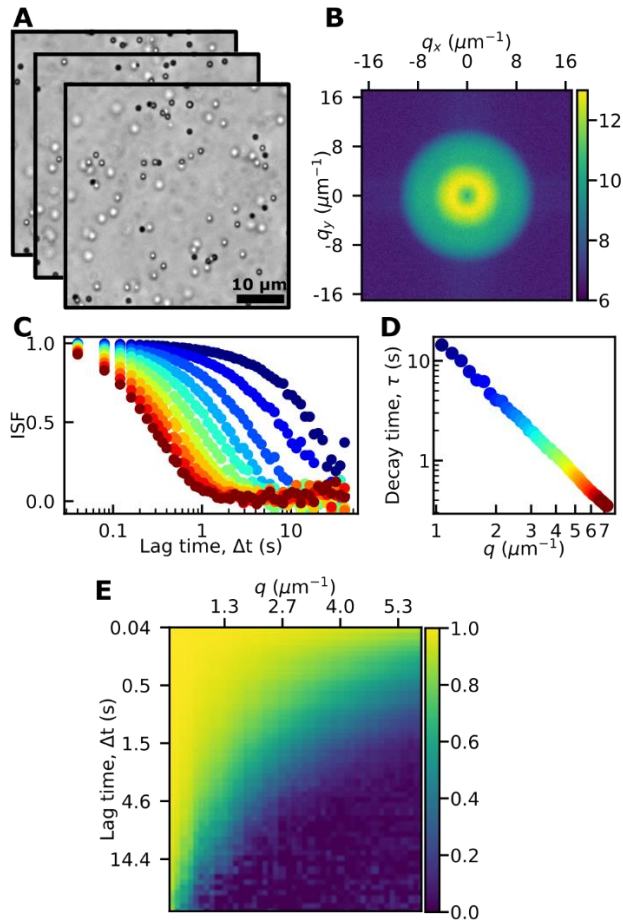

**Figure 1**. DDM quantifies dynamics of microspheres from microscopy movies. (A) Brightfield microscopy images of 1-μm-diameter colloidal particles in a solution of dextran are shown. A sequence of images is captured to perform DDM analysis. (B) From the movie, we calculate the DDM matrix $D(q_x, q_y, \Delta t)$ where $q_x$ and $q_y$ are the x- and y-components of the wave vector and $\Delta t$ is the lag time. The DDM matrix shown is for $\Delta t = 2.64$ s. (C) Given the symmetry of the DDM matrix (expected given that the diffusive dynamics of the colloidal particles are isotropic), we average azimuthally to acquire $D(q, \Delta t)$ where $q$ is the magnitude of the wave vector. Assuming that the DDM matrix can be described by Eqn 2, we use estimates of the amplitude ($A$) and background ($B$) terms to compute the intermediate scattering function (ISF). The ISFs plotted as a function of lag time are for wave vector magnitudes that span from 0.7 to 7.3 μm$^{-1}$. At larger wave vectors (corresponding to smaller length scales and redder colors), the ISFs decay more quickly than at smaller wave vectors (bluer colors). (D) We fit the ISFs to exponential functions to determine a characteristic decay time, $\tau$. A plot of $\tau$ vs $q$ on a log-log scale reveals a power law behavior. In this case, $\tau = (Dq^2)^{-1}$ where $D$ is the diffusion coefficient. (E) In the DDM literature, ISFs are most often depicted as plots of the ISF vs lag time at various wave vectors (as in (C)). However, one could also represent the ISF as a 2D matrix as shown. It is this representation of the ISF which we denoise. Note that lag times are logarithmically spaced on the vertical axis.

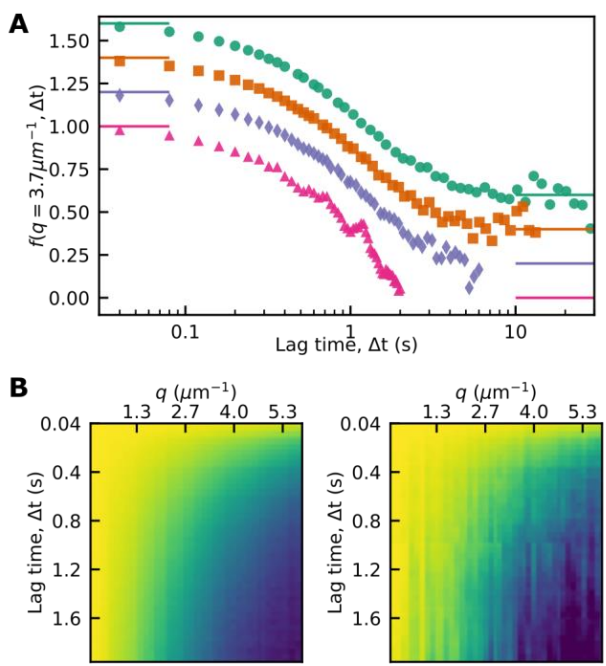

**Figure 2**. DDM analysis on movies with fewer frames results in noisier ISFs. (A) Four ISFs are plotted as a function of lag time, all for the same magnitude of the wave vector, $q$ = 3.73 µm$^{-1}$. These four ISFs are calculated from the same movie of 1 µm beads recorded at 25 frames per second. Curves have been offset to prevent their overlap with the solid lines at the small and large lag times indicating 1 and 0 for each ISF. The top curve (cyan) is the ISF when all 3,000 frames were used for calculating the DDM matrix. The next curve (orange) was generated using only the first 1,000 frames. The following (purple and pink) were generated using the first 500 frames and 51 frames, respectively. (B) The ISFs are depicted as matrices with the vertical axis being lag time and the horizontal axis being the magnitude of the wave vector, $q$. Here, the lag times are linearly spaced from 1 frame to 50 frames (0.04 s to 2 s). The left ISF was generated using the full 3,000 frame movie of diffusing micron-sized beads. The right ISF, which is visibly noisier, was generated using only the first 51 frames of that same movie.

## 3 Results and discussion
### 3.1 Data collection and processing
In all our presented data, we used samples containing 1-µm-diameter colloidal spheres (Fluoresbrite YG Carboxylate Microspheres, Polysciences). We used two optical microscopes, both Olympus IX73 systems where we used a 40× objective, either a LUCPlanFLN 0.60 NA or a PlanN 0.65 NA. The primary difference between the two systems was the camera: either a Hamamatsu Orca-Flash 2.8 CMOS camera or pco.edge 4.2 LT CMOS camera. For data processing, we used in-house DDM software written in Python, PyDDM.[57] For the CNN-ED model, we used the code developed for denoising X-ray photon correlation spectroscopy data which was implemented with the PyTorch framework.[30] We note that while the authors of Ref. 30 denoised a two-time correlation function, our adaptation denoises a 2D matrix representation of the ISF as shown in Fig. 2B.

For training and implementing the CNN-ED model, we used ISFs calculated with DDM. To retrieve the ISF from the DDM matrix, we require a measure or estimate of the amplitude, $A(q)$, and background, $B$, parameters. For these estimates, we assumed $B$ is independent of the wave vector. We found $A(q)$ and $B$ by computing the mean of the squared Fourier-transformed images, $\langle |\tilde{I}(\boldsymbol{q},t)|^2 \rangle_t$, and assumed it equal to $(A(q)+B)/2$.[33, 58] We further assumed that $A(q)$ goes to zero at large $q$.

To determine the diffusion coefficient of the micron-sized beads, either from the ISFs calculated directly from the acquired movies or from the denoised ISFs that are outputted by the CNN-ED model, we first fitted the 1D ISF at each value of $q$ as a function of the lag time to an exponential function of the form $f(q,\Delta t) = \exp(-(\Delta t/\tau(q))^{\gamma(q)})$ where $\gamma(q)$ is a stretching exponent included in our function to account for possible polydispersity or heterogeneity in the dynamics. With the assumption that the dynamics are diffusive, we found the diffusion coefficient, $D$, by taking the mean of $(\tau q^2)^{-1}$ across a range of $q$ values, generally 3.0 to 4.2 µm$^{-1}$. We note that it is more common in DDM studies to estimate $D$ from fitting data to the relationship $\tau(q) = (Dq^2)^{-1}$. When plotting the diffusion coefficient, we used the standard deviation across this range as the error bar.

### 3.2 Training

In our demonstrations of the CNN-ED model with DDM, we performed imaging of 1-µm-diameter colloidal spheres moving diffusively. Therefore, we train this model using images of these colloidal particles suspended in various dextran (500kDa, Fisher BP1580) concentrations. All microscopy data used for training was acquired using the brightfield imaging modality on an Olympus IX73 microscope using a 40× 0.65 NA objective.

Our machine learning approach requires training the CNN-ED model with pairs of 2D matrices: a noisy or "raw" 2D ISF and a corresponding "target" 2D ISF that is significantly less noisy. For training, we generate a target ISF from a 3,000-frame video recorded at 25 frames per second with a resolution of 128×128 pixels. We compute this target ISF using all 3,000 frames with time lags linearly spaced between 1 frame and 50 frames. Thus, a target ISF with dimensions of 50×63 is generated with the first dimension being the lag time and the second being the wavenumber. Given our image resolution of 128×128 and the pixel size of 0.36 µm, our wave vector with the greatest magnitude is 8.6 µm$^{-1}$. However, wave vectors beyond approximately $2\pi\,\mathrm{NA}/\lambda \sim 6.8$ µm$^{-1}$, where NA is the objective numerical aperture and $\lambda$ is the wavelength of light, exceed the theoretical maximum for our spatial resolution. Therefore, we crop the data at large $q$ from our ISFs to generate 50×45 matrices. After this cropping, our highest $q$ is 6.1 µm$^{-1}$. To generate the raw, noisier ISFs, we divide each 3,000-frame movie into chunks of 51 consecutive frames, with chunks overlapping by 25 frames. From each subset of 51 frames, we compute the ISF with, again, lag times linearly spaced between 1 and 50 frames (corresponding to 0.04 s to 2.0 s) and $q$ values that go to 6.1 µm$^{-1}$. Therefore, from a 3,000-frame movie, we end up with 119 raw ISFs. From a single 3,000 frame movie, each of those 119 raw ISFs is paired with the same target ISF.

To generate the ISFs, whether raw or target, from our DDM analysis, we calculated $\langle |\tilde{I}(\boldsymbol{q},t)|^2 \rangle_t$ to determine $A(q)$ and $B$, as described previously. We always calculated $\langle |\tilde{I}(\boldsymbol{q},t)|^2 \rangle_t$ using the full 3,000-frame movie (rather than from the smaller chunks of 51 frames). This approach for determining $A(q)$ and $B$ was also used on the test cases presented here. That is, in the demonstrations of our denoising approach to be shown, when we denoise the ISFs from short chunks of image frames, those ISFs were generated using $A(q)$ and $B$ calculated from much longer videos. This approach relies on the

assumption that $A(q)$ and $B$ are not changing in time, even if the sample's dynamics are not in a steady state.

For training a CNN-ED model appropriate for dynamics that could span a range of timescales, we recorded movies of the 1-μm-diameter colloidal spheres in solutions of various viscosities. We prepared 12 solutions of dextran with concentrations ranging from 0.5% to 9% (w/v). The resulting diffusivities of our microspheres ranged from 0.017 μm$^2$/s to 0.33 μm$^2$/s (or, 0.02 pixel$^2$/frame to 0.39 pixel$^2$/frame). For each of the 12 dextran concentrations, we recorded 10 videos, each of 3,000 frames and with 128×128 pixel resolution. Of this, we split the data into training and validation sets. Seven of the 10 movies at each dextran concentration were used for training and the other three were used for validation. Thus, 252,000 image frames (or 3,000×7×12 where 12 is number of dextran concentrations) which generates 84 target ISFs and 9,996 raw ISFs were used for training. Reserved for validation were 4,284 raw ISFs (357 raw ISFs for each dextran concentration).

We used the CNN-ED model architecture and adapted the code from the work of Konstantinova et al.[30] which used the model to denoise correlation functions generated with X-ray photon correlation spectroscopy. We refer readers to that publication for details of the model and published code. In brief, the CNN encoder consists of two 10-channel convolutional layers with 1×1 kernels. The output of each channel goes through a rectified linear unit (ReLU) activation function. A linear transformation then takes the output to a lower dimensional latent space. We used a latent space of size 16. The decoder is symmetric to the encoder and converts the data from latent space to the ISFs of dimensions 50×45. As the cost function, we simply use the mean squared error between the output of the CNN-ED and the target. A simplified picture of the CNN-ED model layout is shown in Fig. 3A where the raw ISF is reduced to a lower dimensional latent space and then decoded. Because much of the noise in the raw ISF cannot be described with the limited number of variables available in the latent space, the output of the model is less noisy.

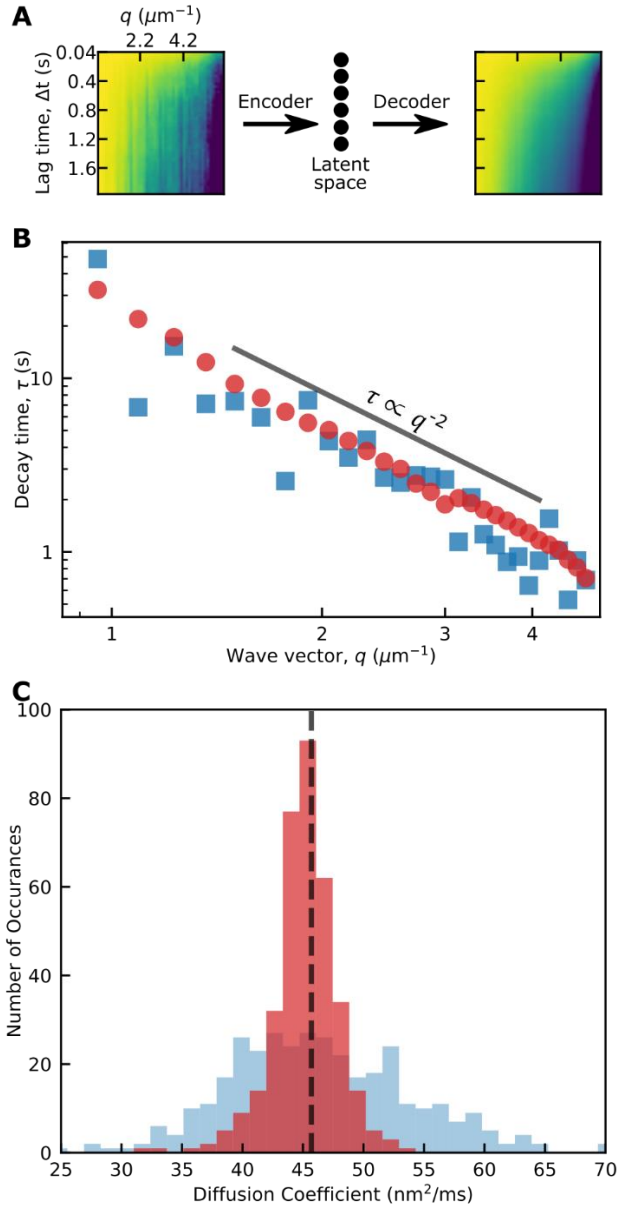

**Figure 3**. Convolutional neural network used to denoise the intermediate scattering function. (A) A representation of the encoder-decoder model is shown. The ISF, seen here with the lag time along the vertical axis and the magnitude of the wave vector along the horizontal axis and with a color map as in Fig. 1E, is encoded to a lower dimensional latent space using 2D convolutional layers and rectified linear unit activation functions. From this latent space, a denoised ISF is then decoded. (B) To quantify the dynamics with DDM, we determine characteristic decay times for each value of $q$ by fitting the ISF to an exponential function. The blue square data points are from an ISF calculated using 51 frames of a movie of 1 µm beads recorded at 25 frames per second. Using the CNN-ED model, that ISF is denoised and used to generate the data points in red. The solid dark gray line shows the power law relationship $\tau \propto q^{-2}$ expected for diffusive dynamics. (C) From the $\tau$ vs $q$ plot, we determine the diffusion coefficient according to $D = (\tau q^2)^{-1}$. This histogram shows the diffusion coefficients determined from ISFs calculated using 51 frames. For three movies of 1 µm beads, each of 3,000 frames, we generate 119 groups of 51 consecutive frames each (thus, each group overlaps by 25 frames). The light blue histogram shows what we determine for the diffusion coefficient of the micron-sized bead from these ISFs. We then run those ISFs through the CNN-ED model for denoising and determine the diffusion coefficient. Those are shown in the red histogram. Note that the three 3,000 frame movies were not used to train the CNN-ED model but were instead reserved for validation.

As a first test of this approach of using a CNN-ED model to denoise DDM-acquired ISFs, we used the movies of micron-sized beads diffusing in dextran that were set aside as the validation data sets. These movies were not used to build the CNN-ED model but they were taken using the same sample and microscopy settings as the data used to build the model. We plotted the determined decay time, $\tau$, as a function of $q$ using an ISF determined from a group of 51 consecutive frames as part of the validation movies recorded as previously described, as shown in Fig. 3B. The data points plotted with blue squares are from the raw ISFs and exhibit noticeable fluctuations from the power law relationship expected for simple diffusive motion. This is as expected given that with only 51 frames we covered a limited range of lag times from 0.04 to 2 s and have few pairs of images to compute the ISF compared to the number of image pairs typically used for DDM. The data points plotted with red circles were generated after passing the same ISF through the trained CNN-ED model. As observed, this $\tau(q)$ data follows the expected power law relationship much better. From the determined $\tau(q)$ and the relationship for diffusive motion, $\tau = (Dq^2)^{-1}$, we found the diffusion coefficient of the beads in the three 3,000-frame validation datasets with a dextran concentration of 5%. Given that we broke each movie into chunks of 51 frames (overlapping by 25 frames), we have $119 \times 3 = 357$ ISFs calculated. The distribution of diffusion coefficients found from these 357 ISFs is shown with the blue histogram in Fig. 3C. In red, we show the distribution of diffusion coefficients after employing the CNN-ED to denoise each ISF. Both distributions are centered around the diffusion coefficient found when DDM was used on all 3,000 frames of each movie, 45.7 nm$^2$/ms. However, when using the CNN-ED model, the distribution of diffusion coefficients has a width that is about 5 times smaller than when using the raw ISFs.

### 3.3 Solution undergoing gelation

As a demonstration of how this machine learning approach to DDM analysis can be useful for studying non-equilibrium samples with dynamics that evolve in time we investigated a protein gel system. We prepared a solution of 6% (w/w) sodium caseinate solution in deionized water containing a trace amount of the same colloidal microspheres used in our model training and validation data. To induce gelation, we added glucono-δ-lactone (GDL) at a concentration of 2.6% (w/w). The GDL causes the pH to progressively lower. Prior studies have shown that as the pH gets near and then goes below the isoelectric point of casein (pH ~ 4.6), the casein particles aggregate and the system gels.[59] For our samples prepared thusly, we found that the system gels in ~45 minutes after adding the GDL as inferred by the apparent immobilization of the colloidal microspheres.

As with the data used for training the CNN-ED model, we acquired images with a 40× objective at a rate of 25 frames per second. However, we used fluorescence imaging rather than the brightfield modality used in the training datasets. Given that the ISFs depend solely on the dynamics and not on the structure or intensity profile of the diffusing objects, we expected the CNN-ED model to work when using this fluorescence modality. An example image of this system is shown in Fig. 4A where we also display particle trajectories, determined through single particle tracking using the trackpy package,[60] of these beads over two 16-second intervals before the sample has fully gelled (bottom left) and after the sample has nearly completely gelled (bottom right).

We also highlight the non-stationary dynamics of this system by displaying the time dependent ISF in Fig. 4B. This two-time correlation function (2TCF) representation is shown for a particular value of $q$ and with the horizontal and vertical axes both being time. Near this function's diagonal that runs from the bottom left to the top right, we have pairs of frames closely separated in time where the ISF would be near one (redder colors). Moving away from that diagonal, the correlation between images decreases (i.e., the ISF decreases to zero). Notice that the red (or more highly correlated) region of the 2TCF broadens as time moves forward which indicates that the dynamics are slowing over time, as expected for this gelling sample. A typical DDM approach would be to average this 2TCF over time (i.e., average together all data points that lie on lines parallel to the bottom left to top right diagonal) or to do so at least over time spans that encompass hundreds or thousands of image frames. In this case, such an approach would mask how the dynamics are evolving in time.

Fig. 4C shows how the ISFs for particular values of $q$ are affected by the number of frames used in DDM. In Fig. 4C(i), we show ISFs computed using 1,000 frames, the lighter gray data points being from a group of 1,000 frames earlier in the gelation process and the black data points from a non-overlapping group of 1,000 frames later in the gelation process, hence the slower decay. The 1,000-frame groups for these ISFs are shown as the boxes with dashed outlines in the 2TCF in Fig. 4B. We notice that the ISFs generated when using 1,000 frames are fairly smooth with lag time. But they are also smooth in $q$. We show this by having plotted the ISFs in Fig. 4C at three different values of $q$. In Fig. 4C(i), it may not be immediately noticeable by eye that there are three sets of points for the light gray and black data since those curves largely overlap. These three data sets are at consecutive values of $q$ (2.12, 2.19, and 2.27 $\mu m^{-1}$). In Fig. 4C(ii), we have plotted ISFs that were calculated using only 51 frames, again at different times in the gelation process (light gray at earlier times and black at later times, as shown with the boxes with the solid outlines in the 2TCF in Fig. 4B). This reduces the maximum lag time accessible. It also significantly increases the noise. We observe that the ISFs are less smooth with lag time and with $q$. The same data from Fig. 4C(ii) is plotted in Fig. 4C(iii) after the ISFs were denoised using the CNN-ED. While the data is not as smooth (in neither lag time nor in $q$) as the ISF generated using 1,000 frames, it is considerably less noisy than the raw ISFs.

In Fig. 4D we plot the diffusion coefficient, $D$, of the micron-sized beads as a function of time. The black dots indicate the diffusion coefficients found by employing standard DDM on non-overlapping groups of 750 frames (spanning a time window of 30s). We also found the diffusion coefficients by using groups of 51 frames, as plotted with the blue data points. While the temporal resolution is finer, there is both enhanced noise in the values of $D$ and a departure from the values of $D$ expected from the DDM analysis over larger time windows (the black data points). In red, we plot our results after using the CNN-ED model to denoise the ISFs generated from groups of 51 frames. There is both less uncertainty in the determined values of $D$ and they are generally closer to what we expect from the analysis of 750 frames (at least during times when the dynamics are relatively stationary, i.e., at times earlier than 90 s).

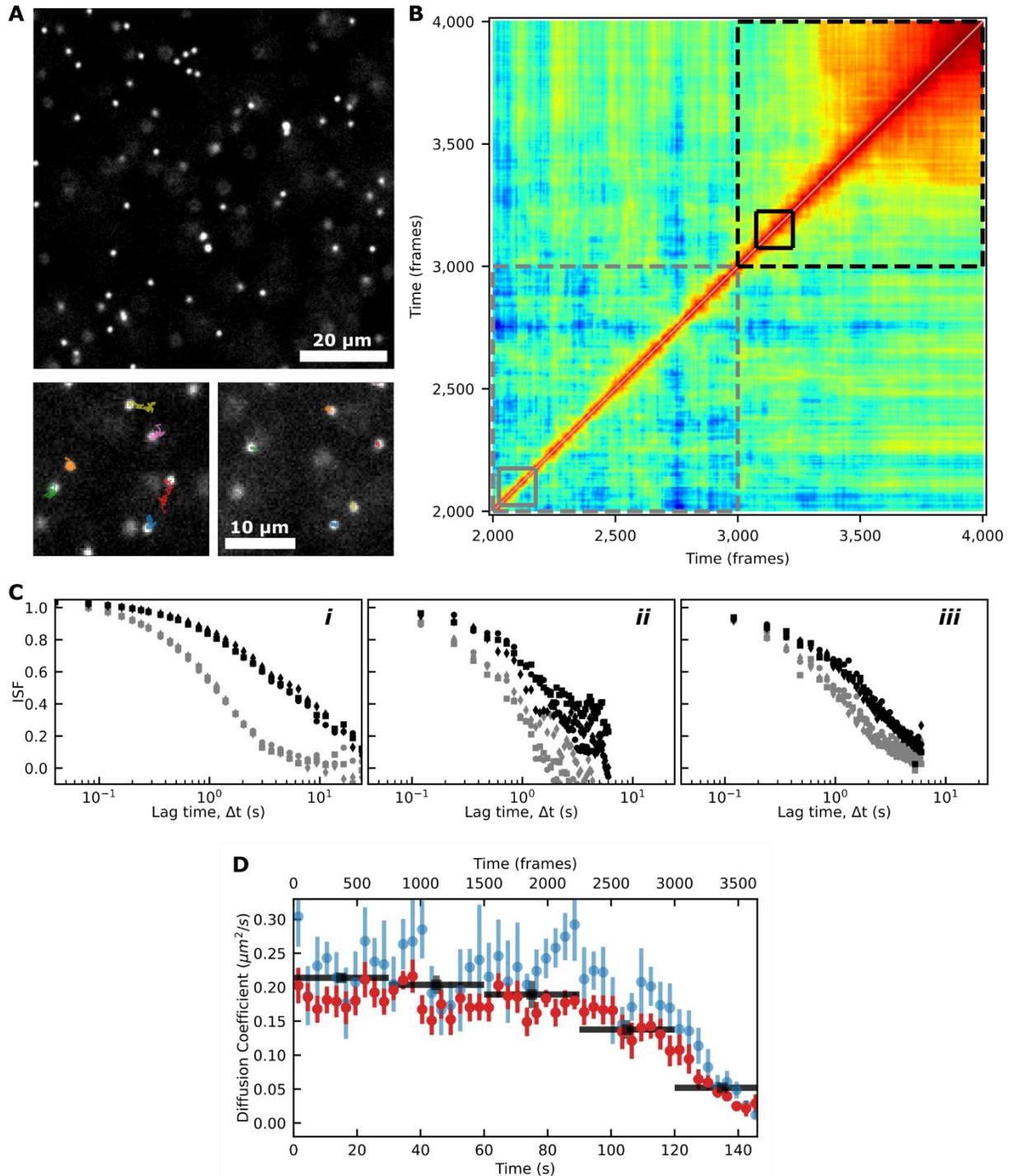

**Figure 4**. Denoising ISFs improves accuracy of the quantification of colloidal dynamics in a non-equilibrium gelling sample. (A) We used fluorescence microscopy to image 1 μm beads in a solution containing sodium caseinate and GDL. We recorded a movie of 4,000 frames at 25 Hz. The bottom two images depict trajectories of particles over a time span of 16 s (400 frames). Left: trajectories shown for six beads from frames 2751 to 3150 (corresponding to t = 110 s to 126 s). Right: trajectories for five beads from frames 3801 to 4200 (corresponding to t = 152 s to 168 s). At the earlier time, the beads are able to move greater distances than compared with the later time due to the casein forming a gel. (B) This depiction of the ISF where instead of computing the ISF at a given lag time, we compute it as a function of two times is referred to as a two-time correlation function (2TCF). Red colors correspond to a high degree of correlation between image frames at the two times (i.e., ISF near 1) while blue colors correspond to a little correlation (i.e., ISF near 0). This depiction of the ISF demonstrates the non-steady state nature of the dynamics. The regions boxed by gray and black solid and dashed lines correspond to the time-averaged ISFs plotted below. (C) Plots of the ISF are shown vs lag time, all for the same three values of $q$, 2.12, 2.19, and 2.27 μm$^{-1}$. Left, *i*: the two ISFs are each computed from 1,000 frames of the recorded movie. The light gray curves show the ISFs using frames 2,000 - 3,000 (the region of the 2TCF outlined by the dashed gray line). The black curves show the ISFs using frames 3,000 - 4,000 (the region of the 2TCF outlined by the dashed black line). Note that for both the light gray and the black data points, there are three sets of points corresponding to three values of $q$. The three curves largely overlap as the values of $q$ are close. As gelation of casein occurred during the recording of the movie, the ISFs calculated from the later frames decay more slowly. Middle, *ii*: the ISFs are computed from 51 frames of the recorded movie. The light gray curves show the ISFs using frames 2,025 - 2,076 (the region of the 2TCF outlined by the solid gray line). The black curves show the ISFs using frames 3,075 - 3,126 (the region of the 2TCF outlined by the solid black line). As with the left plot, we show the ISFs for three different $q$ values. Right, *iii*: the same ISFs are plotted as in the middle plot but after being denoised with our CNN-ED. (D) The diffusion coefficients of the micron-sized beads are plotted as a function of time. The black data points correspond to conventional DDM applied to non-overlapping groups of 750 frames (covering time intervals of 30 s). The blue data points correspond to using groups of 51 frames. The red data points correspond to our results after applying the CNN-ED denoising method on the same data used for generating the blue data points.

### 3.4 On-the-fly scanning

As our next demonstration, we performed on-the-fly scanning of a sample having a gradient in dextran concentration (and, hence, a gradient in the viscosity). This gradient was created by pipetting, from either end of a sample chamber made from a glass slide and coverslip, a 1% dextran solution and a 10% dextran solution, both containing 1-μm-diameter tracer particles. Such a sample was prepared to demonstrate how we can use DDM to quantify the diffusion of tracer particles (for the purpose of microrheology, for example) in a situation where the dynamics will be heterogeneous or non-uniform throughout the sample volume. In such a situation, one approach to quantitatively mapping out the diffusivity of the tracers throughout the sample using DDM would be to record sequences of images at different positions within the sample. This approach of moving the slide to image at a certain location, recording a movie, moving the slide to a new location, and repeating would, of course, be time consuming. Furthermore, if the spatially heterogeneous dynamics were also changing in time, then a protracted measurement method would confound the results.

We investigated an approach to mapping out the dynamics over the sample volume that consists of recording a movie as the slide is moved at a constant velocity with a motorized microscope stage. This on-the-fly scanning method could potentially allow for fast mapping or for high-throughput screens. Two main difficulties to overcome with this approach were (i) the apparent drift velocity of the tracer particles due to the sample scanning needing to be subtracted from the measured dynamics (in our case, to reveal the underlying diffusive motion) and (ii) the spatial resolution of the dynamics mapped out throughout the sample needing to be balanced by the time needed to acquire enough images to accurately capture the dynamics. Naturally, our approach to (ii) was to use the CNN-ED model to denoise the ISFs and quantify the dynamics with a limited number of frames. For (i), we used a previously described extension of DDM that allows for a constant drift velocity, $\boldsymbol{v}$, to be computationally removed. This technique is known as phase differential microscopy ($\varphi$DM) due to the fact that a translation in real space results in a phase shift in Fourier space.[48, 49] Essentially, before subtracting the Fourier transforms of two images separated by a lag time, $\Delta t$, we multiply one of the Fourier transforms by $\exp(-i\phi)$ where $\phi = \boldsymbol{q} \cdot \boldsymbol{v} \Delta t$.

The experimental setup we used is depicted in Fig. 5A. We imaged 1 μm beads with a 40× objective using brightfield microscopy. The sample was on a motorized x-y stage (MS-2000 from Applied Scientific Instrumentation with standard 6.35 mm pitch lead screw) and translated at a constant 1.74 μm/s in the direction along the gradient in dextran concentration. We recorded images at 25 Hz for a total of 20,000 frames (thus, lasting 13.3 minutes and covering a total distance of 1.4 mm). In Fig. 5B we show ISFs calculated from groups of 2,000 non-overlapping frames, all at the same value of $q$. In Fig. 5B(i), we have not corrected for the sample stage velocity, thus we observe oscillations in the ISF, indicative of ballistic motion. Furthermore, the ISFs shown in Fig. 5B(i), captured from different regions of the sample as shown with the colorbar in Fig. 5A, do not show much difference even though the bead diffusivity is slowing as the stage moves toward the higher dextran concentration. This is due to the drift velocity dominating the dynamics regardless of the diffusivity. In Fig. 5B(ii), we have computationally removed the drift using $\varphi$DM. Here, we observe ISFs that resemble exponential decays (rather than showing oscillations) and we note that as the stage position moves forward, the ISFs decay more slowly, reflecting the increased dextran concentration.

Increasing the spatial resolution of our diffusivity mapping would require decreasing the number of frames used for DDM (or, alternatively, moving the stage at a slower speed). If instead of using 2,000 non-overlapping sets of frames, we used 51 frames, then we find the ISFs shown at nine different positions along the gradient in Fig. 5C(i). Here, we have used $\varphi$DM for drift velocity removal. We note that the ISFs are considerably noisier than those in Fig. 5B(ii) in addition to extending to a smaller maximum lag time. By using the CNN-ED model, we can reduce much of the noise in the ISFs as shown in Fig. 5C(ii). Our results from using DDM, with $\varphi$DM in all cases, to find the diffusion coefficients of the micron-sized beads are shown in Fig. 5D when using groups of 2,000 frames (black squares), groups of 51 frames (blue circles), or groups of 51 frames with the CNN-ED model used for denoising (red circles). There are still fluctuations in the measured diffusion coefficients when using the CNN-ED model, but considerably smaller than without the denoising step.

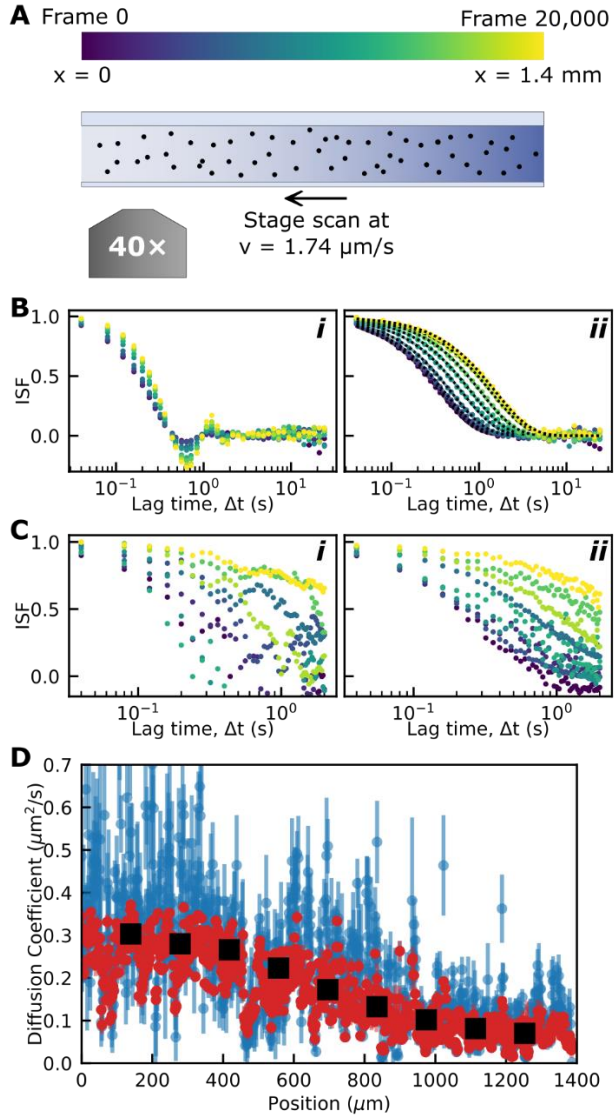

**Figure 5**. Denoising ISFs and φDM were used to perform on-the-fly scanning for fast characterization of a spatial gradient in the diffusivity of micron-sized beads. (A) This schematic of our experiment depicts beads within a solution having a gradient in the dextran concentration. The slide was scanned at 1.74 µm/s along the gradient while images were recorded at 25 frames per second. The top color bar denotes that we recorded 20,000 frames which corresponds to a scanned distance of 1.4 mm. (B) For each plot, nine ISFs are plotted as a function of lag time. All are for the same value of $q$, 3.07 µm$^{-1}$. The nine sets of points are the ISFs computed from non-overlapping groups of 2,000 frames (corresponding to a stage displacement of 139.2 µm) of the entire movie. The nine sets of ISFs are color coded according to the color bar in (A). In (i), the ISFs computed for different times (corresponding to different distances along the dextran gradient) all nearly overlap. We also notice oscillations in the ISFs. These features are because DDM is picking up on the dominant feature of the dynamics: the steady drift velocity of the beads. This drift velocity is the same over the entire movie (due to the stage moving at a constant velocity) so the ISFs measured in regions of the sample with different dextran concentrations are relatively similar. In (ii), we have computationally removed the drift velocity using φDM. We observe that the ISFs (all for the same $q$ = 3.07 µm$^{-1}$) steadily decay more and more slowly as the movie progresses (and, therefore, the region of the sample imaged contains higher concentrations of dextran). (C) In (i), we use groups of 51 frames rather than 2,000 frames to compute the ISFs, using φDM for removal of the drift velocity. As before, these nine ISFs are for the same value of $q$. In (ii), we have used the CNN-ED to denoise the ISFs computed with 51 frames. (D) We plot the diffusion coefficient of the beads as a function

of the position along the dextran gradient. For all data shown here, we have used $\varphi$DM to remove the drift velocity. The black points are from DDM analysis of groups of 2,000 frames. The blue data points are from DDM analysis of groups of 51 frames. The red data points are also from analysis with 51 frames but with the CNN-ED used to denoise the ISFs.

## 4 Conclusions

We have demonstrated how a machine learning approach can assist DDM analysis. Conventionally, DDM is used on movies of several hundred or thousands of frames. This need for a large number of frames can be time consuming and can present problems when non-stationary dynamics are present. Unfortunately, with small numbers of frames used to calculate the DDM matrix (and the ISF), the results are frequently noisy. We have shown that a neural network can be used to denoise the ISFs computed using DDM. This is not an unexpected result given the prior use of neural networks for denoising applications. The test cases we have described show how the use of a CNN-ED model can be used for quantifying the time-dependent diffusivity of tracer beads in a gelling sample and for quantifying the position-dependent diffusivity of tracer beads in a sample having a gradient in viscosity.

In these test cases that we have described, we assumed that the parameters $A(q)$ and $B$ were stationary over time, even though the dynamics, quantified through the parameter $\tau(q)$, was not. There are many soft matter systems of interest where not just the dynamics but also the structure is evolving in time. For example, samples exhibiting photobleaching, changes in opacity, or coarsening[61] would show changes in $A(q)$. If $A(q)$ varied more slowly than $\tau(q)$, then our presented approach could be extended to account for that by determining $A(q)$ using some subset of frames smaller than the total length of the video (as we did with the data presented here) but larger than the interval used to generate the ISFs which get denoised. However, a less straightforward approach would be necessary to deal with situations involving $A(q)$ and $\tau(q)$ evolving at similar rates.

A related limitation to our approach is that it assumes that $A(q)$ and $B$ can be estimated from the images rather than determined through fits to Eqn. 2. Our approach to estimating $A(q)$ and $B$ assumes that there is sufficient contrast between the sample and the background. This condition is easily met with brightly labeled fluorescent beads or with colloidal tracer particles that are highly scattering. However, this approach to estimating $A(q)$ and $B$ may fail when imaging much smaller colloidal particles or objects with less contrast.[31, 62] As with scenarios involving a non-stationary $A(q)$, alternative approaches for denoising DDM results are needed for samples with very little contrast.

Another shortcoming of our method could be situations where the dynamics abruptly shift or rapid decorrelation events occur, rather than the relatively gentle change in dynamics for the scenarios we have investigated here. If those decorrelation events happen over time scales smaller than those used for calculating the to-be-denoised ISFs, then one could miss those events. One approach could be to generate the two-time correlation function, such as that shown in Fig. 4B, to estimate the time scale over which the dynamics are changing and to determine what window size would be appropriate to calculate the ISFs. Such an approach could allow for DDM to be used to study samples exhibiting intermittent decorrelation events. Previous work has explored using time-dependent DDM analysis to study samples where intermittent events occur.[5, 63] If our CNN-ED denoising method were applied to such samples, one would have to test that rapid decorrelation events are not removed along with noise and would likely have to explore the appropriate time window for calculating ISFs such that the dynamics can be accurately measured without losing information about the intermittent events.

We note that the data we denoised with a CNN-ED model were all of diffusive dynamics that fell within the range of diffusivities of the data sets used for model training. We have yet to explore how well this model would work for quantifying dynamics that are faster or slower than that captured in the training data. To build up a library of training data to create a CNN-ED model that is likely to work in a wider range of scenarios we are exploring the use of computer simulations. Additionally, to improve this method of denoising DDM data, it will be important to understand how well a CNN-ED model would perform, based on various kinds of training data, in situations with more complex dynamics. For example, an interesting case would be a sample where the decay time, $\tau(q)$, is not a monotonic function of $q$. Such a situation is possible in polydisperse colloidal suspensions where, due to the particles' different structure functions, the contributions to the DDM signal of differently sized particles will depend on $q$.[34] Other classes of dynamics that have been investigated with DDM were mentioned in the Introduction and to what degree our model would fail when presented with more complex dynamics is worth exploring in future work.

There are numerous further avenues to explore how machine learning could enhance DDM beyond addressing the shortcomings described above. Firstly, we limited ourselves to employing the CNN-ED architecture used in prior work.[30] It would be interesting to investigate other neural network architectures. Moreover, one could try other denoising algorithms. For example, in the previous work on using a CNN-ED model to denoise X-ray photon correlation spectroscopy data,[30] denoising methods such as using Gaussian filters or linear principle components-based filters were tested. These methods did not perform as well as the CNN-ED model in that work (and we would expect similar results in our use of CNN-ED models on DDM data), but there may be DDM applications of these more easy-to-use image denoising filters that are worth exploring. Secondly, we denoised the 2D ISFs generated from DDM. Other approaches would be to denoise 3D datasets (such as the 2D ISFs we used with an added third dimension being time) or 2TCF. In particular, it may be helpful to denoise the 3D 2TCF (like the one shown in Fig. 4B with the added third dimension being $q$) in cases where we expect the ISF to be smooth in lag time, $q$, and time (which is what we expect for all the data shown here). Thirdly, a more robust method of uncertainty estimation, such as presented in earlier work,[27] would improve DDM workflow and allow for better assessments of the credibility of the denoising steps. As machine learning approaches become more standard and easier to implement, we expect that this will allow DDM to be used in more non-equilibrium and high throughput scenarios.

### Acknowledgements
This project has been made possible in part by grant number 2023-328570 from the Chan Zuckerberg Initiative DAF, an advised fund of Silicon Valley Community Foundation. RM acknowledges support from the Research Corporation for Science Advancement (award no. 27459). GM was supported by an NSF REU Award (CHE 2050846). RM and JS acknowledge support from an NSF DMREF Award (DMR 2119663).

### Author contributions
Conceptualization: R.M.
Formal analysis: G.M., J.S., S.D., D.G., E.K., R.Y., R.M.
Investigation: G.M., J.S., S.D., J.C.A.
Methodology: G.M., J.S., S.D., R.Y., R.M.
Resources: R.M.

Software: G.M., J.S., S.D., D.G., R.M.
Supervision: R.M.
Visualization: G.M., J.S., E.K., R.M.
Writing – original draft: G.M., R.M.
Writing – review & editing: G.M., R.M.

**Conflicts of interest**
There are no conflicts to declare.

**References**
1. R. Cerbino and V. Trappe, *Physical Review Letters*, 2008, **100**, 188102.
2. K. He, M. Spannuth, J. C. Conrad and R. Krishnamoorti, *Soft Matter*, 2012, **8**, 11933-11938.
3. L. G. Wilson, V. A. Martinez, J. Schwarz-Linek, J. Tailleur, G. Bryant, P. N. Pusey and W. C. K. Poon, *Physical Review Letters*, 2011, **106**, 018101.
4. M. Drechsler, F. Giavazzi, R. Cerbino and I. M. Palacios, *Nature Communications*, 2017, **8**, 1-11.
5. Y. Gao, J. Kim and M. E. Helgeson, *Soft Matter*, 2015, **11**, 6360-6370.
6. J. H. Cho, R. Cerbino and I. Bischofberger, *Physical Review Letters*, 2020, **124**, 088005.
7. G. Lee, G. Leech, M. J. Rust, M. Das, R. J. McGorty, J. L. Ross and R. M. Robertson-Anderson, *Science Advances*, 2021, DOI: 10.1126/sciadv.abe4334.
8. M. Al-Shahrani and G. Bryant, *Physical Chemistry Chemical Physics*, 2022, DOI: 10.1039/D2CP02034C.
9. R. Cerbino, F. Giavazzi and M. E. Helgeson, *Journal of Polymer Science*, 2021, **60**, 1079-1089.
10. R. Cerbino and P. Cicuta, *The Journal of Chemical Physics*, 2017, **147**, 110901.
11. F. Giavazzi and R. Cerbino, *Journal of Optics*, 2014, **16**, 083001.
12. K. Zhang, X. Gong and Y. Jiang, *Advanced Functional Materials*, 2024, **n/a**, 2315177.
13. K. Ayush, A. Seth and T. K. Patra, *Soft Matter*, 2023, **19**, 5502-5512.
14. L. E. Altman and D. G. Grier, *Soft Matter*, 2023, **19**, 3002-3014.
15. K. R. Lennon, G. H. McKinley and J. W. Swan, *Proceedings of the National Academy of Sciences*, 2023, **120**, e2304669120.
16. A. Lizano and X. Tang, *Soft Matter*, 2023, **19**, 3450-3457.
17. M. G. Smith, J. Radford, E. Febrianto, J. Ramírez, H. O'Mahony, A. B. Matheson, G. M. Gibson, D. Faccio and M. Tassieri, *AIP Advances*, 2023, **13**, 075315.
18. M. R. Bailey, F. Grillo and L. Isa, *Soft Matter*, 2022, **18**, 7291-7300.
19. P. S. Clegg, *Soft Matter*, 2021, **17**, 3991-4005.
20. E. Bedolla, L. C. Padierna and R. Castañeda-Priego, *Journal of Physics: Condensed Matter*, 2020, **33**, 053001.
21. N. E. Jackson, M. A. Webb and J. J. de Pablo, *Current Opinion in Chemical Engineering*, 2019, **23**, 106-114.
22. Z. Zhou, C. Li, X. Bi, C. Zhang, Y. Huang, J. Zhuang, W. Hua, Z. Dong, L. Zhao, Y. Zhang and Y. Dong, *npj Computational Materials*, 2023, **9**, 1-14.
23. J. M. Ede and R. Beanland, *Ultramicroscopy*, 2019, **202**, 18-25.
24. V. Mannam, Y. Zhang, Y. Zhu, E. Nichols, Q. Wang, V. Sundaresan, S. Zhang, C. Smith, P. W. Bohn and S. S. Howard, *Optica*, 2022, **9**, 335-345.

25. Y. Wang, H. Pinkard, E. Khwaja, S. Zhou, L. Waller and B. Huang, *Optics Express*, 2021, **29**, 41303-41312.
26. E. Xypakis, V. d. Turris, F. Gala, G. Ruocco and M. Leonetti, *Optics Express*, 2023, **31**, 43838-43849.
27. M. Gu, Y. Luo, Y. He, M. E. Helgeson and M. T. Valentine, *Physical Review E*, 2021, **104**, 034610.
28. R. L. Martineau, A. V. Bayles, C.-S. Hung, K. G. Reyes, M. E. Helgeson and M. K. Gupta, *Advanced Biology*, 2022, **6**, 2101070.
29. T. Konstantinova, L. Wiegart, M. Rakitin, A. M. DeGennaro and A. M. Barbour, *Physical Review Research*, 2022, **4**, 033228.
30. T. Konstantinova, L. Wiegart, M. Rakitin, A. M. DeGennaro and A. M. Barbour, *Scientific Reports*, 2021, **11**, 14756.
31. M. S. Safari, M. A. Vorontsova, R. Poling-Skutvik, P. G. Vekilov and J. C. Conrad, *Physical Review E*, 2015, **92**, 042712.
32. M. Reufer, V. A. Martinez, P. Schurtenberger and W. C. K. Poon, *Langmuir*, 2012, **28**, 4618-4624.
33. R. Cerbino, D. Piotti, M. Buscaglia and F. Giavazzi, *Journal of Physics: Condensed Matter*, 2017, **30**, 025901.
34. J. Bradley, V. A. Martinez, J. Arlt, J. Royer and W. Poon, *Soft Matter*, 2023, **19**, 8179-8192.
35. A. Pal, V. A. Martinez, T. H. Ito, J. Arlt, J. J. Crassous, W. C. K. Poon and P. Schurtenberger, *Science Advances*, 2020, **6**, eaaw9733.
36. D. Germain, M. Leocmach and T. Gibaud, *American Journal of Physics*, 2016, **84**, 202-210.
37. O. A. Croze, V. A. Martinez, T. Jakuszeit, D. Dell'Arciprete, W. C. K. Poon and M. A. Bees, *New Journal of Physics*, 2019, **21**, 063012.
38. M. Brizioli, T. Sentjabrskaja, S. U. Egelhaaf, M. Laurati, R. Cerbino and F. Giavazzi, *Frontiers in Physics*, 2022, **10**.
39. J. A. Richards, V. A. Martinez and J. Arlt, *arXiv:2102.11094 [cond-mat]*, 2021.
40. P. J. Lu, F. Giavazzi, T. E. Angelini, E. Zaccarelli, F. Jargstorff, A. B. Schofield, J. N. Wilking, M. B. Romanowsky, D. A. Weitz and R. Cerbino, *Physical Review Letters*, 2012, **108**, 218103.
41. D. M. Wulstein, K. E. Regan, R. M. Robertson-Anderson and R. McGorty, *Optics Express*, 2016, **24**, 20881-20894.
42. A. V. Bayles, T. M. Squires and M. E. Helgeson, *Soft Matter*, 2016, **12**, 2440-2452.
43. R. Nixon-Luke, J. Arlt, W. C. K. Poon, G. Bryant and V. A. Martinez, *Soft Matter*, 2022, DOI: 10.1039/D1SM01598B.
44. F. Giavazzi, D. Brogioli, V. Trappe, T. Bellini and R. Cerbino, *Physical Review E*, 2009, **80**, 031403.
45. P. Edera, D. Bergamini, V. Trappe, F. Giavazzi and R. Cerbino, *Physical Review Materials*, 2017, **1**, 073804.
46. A. V. Bayles, T. M. Squires and M. E. Helgeson, *Rheologica Acta*, 2017, **56**, 863-869.
47. F. Giavazzi, P. Edera, P. J. Lu and R. Cerbino, *The European Physical Journal E*, 2017, **40**, 97.
48. J. A. Richards, V. A. Martinez and J. Arlt, *Soft Matter*, 2021, **17**, 3945-3953.
49. R. Colin, R. Zhang and L. G. Wilson, *Journal of The Royal Society Interface*, 2014, **11**, 20140486.
50. M. Arko and A. Petelin, *Soft Matter*, 2019, **15**, 2791-2797.
51. R. You and R. McGorty, *Review of Scientific Instruments*, 2021, **92**, 023702.
52. S. Aime and L. Cipelletti, *Soft Matter*, 2019, **15**, 213-226.
53. K. R. Peddireddy, R. Clairmont, P. Neill, R. McGorty and R. M. Robertson-Anderson, *Nature Communications*, 2022, **13**, 5180.

54. F. Giavazzi, C. Haro-Pérez and R. Cerbino, *Journal of Physics: Condensed Matter*, 2016, **28**, 195201.
55. M. S. Safari, R. Poling-Skutvik, P. G. Vekilov and J. C. Conrad, *npj Microgravity*, 2017, **3**, 21.
56. V. A. Martinez, R. Besseling, O. A. Croze, J. Tailleur, M. Reufer, J. Schwarz-Linek, L. G. Wilson, M. A. Bees and W. C. K. Poon, *Biophysical Journal*, 2012, **103**, 1637-1647.
57. H. N. Verwei, G. Lee, G. Leech, I. I. Petitjean, G. H. Koenderink, R. M. Robertson-Anderson and R. J. McGorty, *JoVE (Journal of Visualized Experiments)*, 2022, DOI: 10.3791/63931, e63931.
58. F. Giavazzi, C. Malinverno, G. Scita and R. Cerbino, *Frontiers in Physics*, 2018, **6**.
59. M. Leocmach, C. Perge, T. Divoux and S. Manneville, *Physical Review Letters*, 2014, **113**, 038303.
60. D. B. Allan, T. Caswell, N. C. Keim, C. M. van der Wel and R. W. Verweij, *Journal*, 2024.
61. F. Giavazzi, V. Trappe and R. Cerbino, *Journal of Physics: Condensed Matter*, 2020, **33**, 024002.
62. C. Guidolin, C. Heim, N. B. P. Adams, P. Baaske, V. Rondelli, R. Cerbino and F. Giavazzi, *Macromolecules*, 2023, **56**, 8290-8297.
63. R. J. McGorty, C. J. Currie, J. Michel, M. Sasanpour, C. Gunter, K. A. Lindsay, M. J. Rust, P. Katira, M. Das, J. L. Ross and R. M. Robertson-Anderson, *PNAS Nexus*, 2023, **2**, pgad245.